\documentclass{osa-article}

\journal{osajournal}

\articletype{Research Article}

\usepackage{lineno}

\newcommand{\vect}[1]{{\bf #1}}
\renewcommand{\d}{{\rm d}}
\renewcommand{\i}{{\rm i}}
\newcommand{\pard}{{\rm\partial}}
\newcommand{\e}{\mathop{\rm e}\nolimits}

\begin{document}

\title{Light-matter interaction problem in classical and quantum optics}

\author{Yuriy Akimov}

\address{Institute of High Performance Computing, 1 Fusionopolis Way, \#16-16 Connexis, Singapore 138632}

\email{akimov@ihpc.a-star.edu.sg} 



\begin{abstract}
Understanding of light-matter interaction is a keystone in mastering classical and quantum optics. This paper gives an overview of the fundamental principles used in these two fields for description of light-matter interaction. By exploring the simplest type of matter composed of charge-free particles bearing magnetic moments only, differences in the fundamental principles of classical and quantum optics are discussed and clarified.  
\end{abstract}


\section{Introduction}

Mastering optics requires clear understanding of the fundamental principles laid in the basis of different optics domains. It can be reached with direct comparison of the descriptions provided by different theories. In this paper, we will compare approaches used in classical and quantum optics for description of light-matter interaction. As an example, we will consider a charge-free matter, whose dynamics is completely governed by spin-spin interactions of particles. This is the simplest model of matter that enables very clear and concise comparison of the fundamental principles used in the two optics domains. 

In section \ref{sec:Light-matter interaction}, we will remind the general picture of light-matter interaction. The case of a charge-free matter composed of spin-bearing particles will be considered in section \ref{sec:Spin-spin interaction of charge-free particles}, where we will review both classical and quantum descriptions. In section \ref{sec:Quantum optics limitation}, we will discuss the limitations brought by different principles used in the two optics domains for description of light-matter interaction. Summary will be given in section \ref{sec:Conclusions}.      

\section{Light-matter interaction}
\label{sec:Light-matter interaction}

To remind the general picture of light-matter interaction, let us consider two elementary objects 1 and 2, which are characterized by their own {\it localized} charge density $\rho_{1,2}(\vect r,t)$ and current density $\vect J_{1,2}(\vect r,t)$. They can be elementary particles or macroscopic objects. To describe their interaction at a distance, we have to assign additional {\it delocalized} fields to them, as their localized fields $\rho_{1,2}(\vect r,t)$ and $\vect J_{1,2}(\vect r,t)$ can be used for description of direct (impact) interaction only. These additional fields should be able to spread over the entire space and interact with localized fields of other objects. This is how we introduce the concept of electromagnetic fields $\vect E_{1,2}(\vect r,t)$ and $\vect B_{1,2}(\vect r,t)$ through which two objects interact one another. This concept obeys the {\it causality principle}, following which electromagnetic fields $\vect E_{1,2}(\vect r,t)$ and $\vect B_{1,2}(\vect r,t)$ have sources in terms of their respective charges and currents given by $\rho_{1,2}(\vect r,t)$ and $\vect J_{1,2}(\vect r,t)$. 

If we look at this interaction from the mathematics point of view, we can notice that it is the self-consistent interaction of two groups of fields -- delocalized fields 
\begin{eqnarray}
	\vect E(\vect r,t)=\vect E_1(\vect r,t)+\vect E_2(\vect r,t),\label{Eq:E}\\
	\vect B(\vect r,t)=\vect B_1(\vect r,t)+\vect B_2(\vect r,t),\label{Eq:B}
\end{eqnarray}
which are called as light, and localized fields 
\begin{eqnarray}
	\rho(\vect r,t)=\rho_1(\vect r,t)+\rho_2(\vect r,t),\\
	\vect J(\vect r,t)=\vect J_1(\vect r,t)+\vect J_2(\vect r,t),
\end{eqnarray}
representing matter. Interaction of these groups of fields, 
\begin{equation}
	\vect E(\vect r,t), \vect B(\vect r,t) \leftrightarrow \rho(\vect r,t), \vect J(\vect r,t),
\end{equation}
is commonly called light-matter interaction. 

Traditionally, electromagnetic fields (\ref{Eq:E}) and (\ref{Eq:B}) are written in terms of the scalar, $\varphi(\vect r,t)$, and vector, $\vect A(\vect r,t)$, potentials \cite{stratton:1941, Landau:1980}, 
\begin{eqnarray}
	&\displaystyle\vect E(\vect r,t)=-\nabla\varphi(\vect r,t)-\frac{\pard\vect A(\vect r,t)}{\pard t},\\
	&\displaystyle\vect B(\vect r,t)=\nabla\times\vect A(\vect r,t).
\end{eqnarray} 
With these potentials, light-matter interaction can be formulated as the interaction 
\begin{equation}
	\varphi(\vect r,t), \vect A(\vect r,t) \leftrightarrow \rho(\vect r,t), \vect J(\vect r,t)
\end{equation}
between the fields $\varphi(\vect r,t), \vect A(\vect r,t)$ from the light side and $\rho(\vect r,t), \vect J(\vect r,t)$ from the matter side. In this interaction, the light fields change distribution of the matter fields, which in their turn modify distribution of the light fields. As a result, we get a closed loop of self-consistent interaction between the two groups of fields. 

\section{Spin-spin interaction of charge-free particles}
\label{sec:Spin-spin interaction of charge-free particles}

To simplify our further discussion of light-matter interaction, we consider the particular case of a charge-free matter, when 
\begin{equation}
	\rho(\vect r,t)\equiv0.\label{Eq:rho}
\end{equation}
This case can be treated within the Coulomb or Lorentz gauge with  
\begin{equation}
	\varphi(\vect r,t)\equiv0. \label{Eq:varphi}
\end{equation} 
Then, the light-matter interaction can be reduced to interaction of two fields only,
\begin{equation}
	\vect A(\vect r,t) \leftrightarrow \vect J(\vect r,t),
\end{equation}
which appear fully transverse, 
\begin{eqnarray}
	&\displaystyle
	\nabla\cdot\vect A(\vect r,t)=0,\label{Eq:A}\\
	&\displaystyle
	\nabla\cdot\vect J(\vect r,t)=0.\label{Eq:J}
\end{eqnarray}
This is the simplest case of light-matter interaction, where the current is composed of intrinsic magnetic moments only and the matter dynamics is governed by spin-spin interactions of charge-free elementary particles. 

\subsection{Classical optics description}

Now, let us discuss how spin-spin interaction of charge-free particles is treated in classical optics. Light dynamics is given by Maxwell's equations, which  reduce under conditions (\ref{Eq:rho})--(\ref{Eq:J}) to the following wave equation \cite{stratton:1941, Landau:1980}:
\begin{equation}
	\nabla^2\vect A(\vect r,\omega)+k_0^2\vect A(\vect r,\omega)=-\mu_0 \vect J(\vect r,\omega),\label{Eq:cl-A}
\end{equation} 
written in the $(\vect r,\omega)$ space with angular frequency $\omega\geq0$. Here, $k_0=\omega/c$ is the free-space wavenumber with $c$ being the speed of light in vacuum.
In the right hand side of Eq.~(\ref{Eq:cl-A}), we have current density $\vect J(\vect r,\omega)$ as the source field for excitation of $\vect A(\vect r,\omega)$. 

As for mater dynamics, we have several classical models based on either Newtonian mechanics or theory of relativity (special or general)\cite{Landau:1980, Landau:1984}. All these models can commonly be represented with a generally nonlinear and nonlocal operator $\hat G_c$ that describes how electromagnetic potential modifies current density, 
\begin{equation}
	\vect J(\vect r,\omega)=\hat G_c\vect A(\vect r,\omega).\label{Eq:cl-J}
\end{equation} 
Without focusing on any particular model of matter dynamics, we can conclude that classical description of particles' spin-spin interaction is self-consistent, as it should be for light-matter interaction: $\vect J(\vect r,\omega)$ excites $\vect A(\vect r,\omega)$ as given by Eq.~(\ref{Eq:cl-A}), and $\vect A(\vect r,\omega)$ modifies $\vect J(\vect r,\omega)$ following Eq.~(\ref{Eq:cl-J}).

\subsection{Quantum optics description}

Next, we consider quantum optics description of spin-spin interaction of charge-free particles. Main differences of quantum approach from classical one are brought by the use of additional concept -- the Heisenberg uncertainty principle. Following this principle, all physical quantities are no longer functions, but operators. As this principle is applied to both light and matter, the vector potential and current density are now operators. The operator $\hat{\vect A}(\vect r,\omega)$ is given in the form of quantized free field \cite{Berestetskii:1982,Feynman:1998}, 
\begin{equation}
	\hat{\vect A}(\vect r,\omega)=\sum\limits_{\alpha=1}^2\int\limits_0^{4\pi}\d\Omega_{\vect n}
	[\hat a_{\alpha\vect n}\vect A_{\alpha\vect n}(\vect r,\omega)+\hat a_{\alpha\vect n}^+\vect A^*_{\alpha\vect n}(\vect r,\omega)],\label{Eq:qu-A}
\end{equation} 
governed by the non-commutative operators of creation $\hat a_{\alpha\vect n}$ and annihilation $\hat a^+_{\alpha\vect n}$ of photons, \begin{equation}
	\hat a_{\alpha\vect n}\hat a_{\alpha\vect n}^+-\hat a_{\alpha\vect n}^+\hat a_{\alpha\vect n}=1,\label{commutation}
\end{equation}
where $\alpha=1,2$ depict the polarization state of photons propagating in the direction specified by the unit vector $\vect n$, and
$\vect A_{\alpha\vect n}(\vect r,\omega)=\vect A_{\alpha\vect n}^{(0)}(\omega)\exp(\i k_0 \vect n\cdot\vect r)$ with $\vect A_{\alpha\vect n}^{(0)}(\omega)$ being the vector potential amplitude normalized to one photon in the volume and featuring $\vect A_{\alpha\vect n}^{(0)}(\omega)\cdot \vect n=0$. Following commutation relation (\ref{commutation}) given by the Heisenberg uncertainty principle, quantum light field can never be measured with absolute precision in contrast to its classical analog that is precisely measurable.

As for matter dynamics, there exist several quantum models based on either non-relativistic Pauli equation or relativistic Dirac equation \cite{Berestetskii:1982,Feynman:1998}. All these approaches describe the quantum effect of electromagnetic fields on matter dynamics and can be written with a generally nonlinear and nonlocal operator $\hat G_q$, similarly to classical models,
\begin{equation}
	\hat{\vect J}(\vect r,\omega)=\hat G_q\hat{\vect A}(\vect r,\omega).\label{Eq:qu-J}
\end{equation} 
If we look at this description from the light-matter interaction point of view, we see how light changes the matter state with Eq.~(\ref{Eq:qu-J}), but matter does not give any feedback to light in Eq.~(\ref{Eq:qu-A}). {\it The quantum optics description of spin-spin interaction of particles appears unable to provide a closed set of equations for self-consistent description of light-matter interaction.} This demonstrates the intrinsic limitation of quantum optics formalism brought by the free field used in the description of light. 
 
\subsection{Free-field model}

Recall, free fields are the special type of sourceless light, which exists independently of matter. These fields exist by themselves and cannot be changed by any physical means. This is the point, where quantum optics violates the causality principle for light, following which, electromagnetic fields are excited by charges and currents and cannot exist independently of them. Furthermore, the use of free fields brings us to the paradox, when {\it quantum version of the light emission problem simply cannot be formulated}, as we do not have any physical means to excite free fields. Also, this makes the control over the state of quantum light generally impossible. Then, how do we change light state in quantum calculations? We do it manually, changing the photon numbers by our hands before solving the quantum equation for matter. We do it in the same way, we manually change state of matter to get the probabilities for all possible  transitions. This similarity with the treatment of matter state gives us the perception of legitimate treatment of light state. If not the causality principle. By changing light state independently of matter one, we destroy the causality in light emission and break the closed loop of light-matter interaction. 

\begin{figure}[t]
\begin{center}
\begin{tabular}{c}
\includegraphics[height=5.5cm]{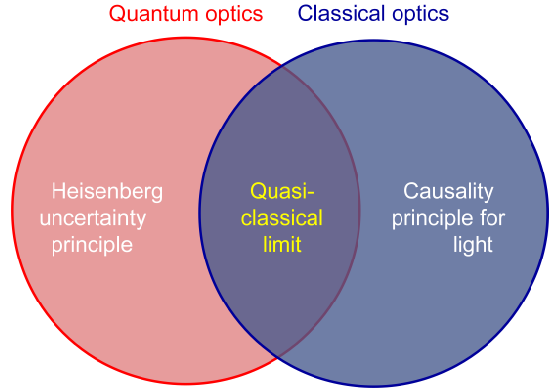}
\end{tabular}
\end{center}
\caption 
{\label{fig:Fig2}
Relationship diagram for classical and quantum optics domains. } 
\end{figure} 

Consideration of spin-spin interaction of particles allows us not only to clarify the fundamental differences in classical and quantum optics, but also to generalize them to other types of light-matter interaction. Since quantum optics deals with free fields only and does not have any other models for light description, its relation to classical optics can be formulated as shown in Fig.~\ref{fig:Fig2}: {\it quantum optics adds up the Heisenberg uncertainty principle to classical optics, but excludes the causality principle for light}. Thus, quantum approach appears unable to completely incorporate classical one for any type of light-matter interaction\cite{Akimov:2021-1,Akimov:2021-2}. Indeed, in the quasi-classical limit of large numbers of photons, we can relax the Heisenberg principle for light and perform the transition in description of light from the operator form to the functional one. But, it never revives the causality principle for light, which is initially excluded in quantum approach with the use of free fields. In other words, {\it quantum optics is not completely consistent with classical description}. To demonstrate this, we return to quantum description of spin-spin interaction of charge-free particles and explore it in the quasi-classical limit.    

\subsection{Quasi-classical limit} 

In the quasi-classical limit of quantum light with large number of photons $N_{\alpha\vect n}(\omega)\gg1$, when the Heisenberg uncertainty is relaxed to $\hat a_{\alpha\vect n}\hat a_{\alpha\vect n}^+\approx\hat a_{\alpha\vect n}^+\hat a_{\alpha\vect n}$, we can rewrite the light field in the functional (precisely measurable) from,
\begin{eqnarray}
	\vect A(\vect r,\omega)=\vect A_{\rm free}(\vect r,\omega)=2\;{\rm Re}\sum\limits_{\alpha=1}^2\int\limits_0^{4\pi}\d\Omega_{\vect n}
	\sqrt{N_{\alpha\vect n}(\omega)}\vect A_{\alpha\vect n}(\vect r,\omega).
\end{eqnarray} 
This is the classical free field given by the homogeneous (sourceless) wave equation, 
\begin{equation}
	\nabla^2\vect A_{\rm free}(\vect r,\omega)+k_0^2\vect A_{\rm free}(\vect r,\omega)=0.
\end{equation} 
In contrast to this field, the field given by classical description obeys the inhomogeneous wave equation with the additional source term $-\mu_0\vect J(\vect r,\omega)$ in the right-hand side [see Eq.~(\ref{Eq:cl-A})]. Following the causality principle, classical fields are given by the forced solution of the wave equation
\begin{eqnarray}
\vect A(\vect r,\omega)=\vect A_{\rm forced}(\vect r,\omega)=\frac{\mu_0}{2\pi}{\rm Re}\int\limits_V\frac{\e^{\i k_0 |\vect r -\vect r'|}}{|\vect r-\vect r'|}\;\vect J(\vect r',\omega)\;\d^3\vect r', \label{A_forced_r}
\end{eqnarray} 
which is quite different from free-field solution $\vect A_{\rm free}(\vect r,\omega)$ of the quasi-classical limit. This can be seen 
from Fourier images of the free and forced solutions written in the $(\vect k,\omega)$ space with $\vect k\in \mathbb{R}^3$, 
\begin{eqnarray}
	&\displaystyle 
	\vect A_{\rm free}(\vect k,\omega)=\int\limits_0^{4\pi}\d\Omega_{\vect n}[\vect A_{\alpha\vect n}(\omega)\delta(k_0 \vect n-\vect k)+\vect A^*_{\alpha\vect n}(\omega)\delta(k_0 \vect n+\vect k)],\label{A_free}
\end{eqnarray} 
\begin{eqnarray}
&\displaystyle \vect A_{\rm forced}(\vect k,\omega)=\frac{\mu_0\vect J(\vect k,\omega)}{k^2-k_0^2},\label{A_forced}
\end{eqnarray} 
where we used
\begin{eqnarray}
	&\displaystyle 
	\vect A_{\alpha\vect n}(\omega)=(2\pi)^{3/2}\sum\limits_{\alpha=1}^2
	\sqrt{N_{\alpha\vect n}(\omega)}\vect A_{\alpha\vect n}^{(0)}(\omega),	
	\\	
	&\displaystyle 
	\vect J(\vect k,\omega)=2(2\pi)^{-3/2}\int\limits_V{\rm Re}[\vect J(\vect r,\omega)]\e^{-\i \vect k\cdot\vect r}\d^3\vect r.
\end{eqnarray} 
According to them, free-field solution $\vect A_{\rm free}(\vect k,\omega)$ is limited to the light cone given by $k=k_0$ and contains only propagating fields, while forced solution $\vect A_{\rm forced}(\vect k,\omega)$ is generally broadband and contains not only propagating fields with $k=k_0$, but also evanescent fields given by Fourier harmonics with $k\neq k_0$. Namely the evanescent fields are the main  difference in free and forced solutions making free fields unsuitable for description of forced fields. However, far away from sources, contribution of evanescent fields decreases, making the propagating Fourier harmonics dominant in the forced solution. This is the case, when the free-field solution gets helpful. 

Indeed, if we do the coordinate transform from the laboratory system of sources to the far-field zone with the origin at $\vect r=\vect R_0$, we can write 
$$\vect r=\vect R_0+\vect R,$$
where $\vect R$ is the new radius vector in the far-field zone system. By the far-field zone, we mean a part of space, located far away from the sources with $R_0\gg r'$. In this zone, we consider a limited domain, whose size is much smaller than the distance from the sources, $R\ll R_0$, but much larger than the wavelength, $R\gg k_0^{-1}$. In this limited far-field domain, forced solution (\ref{A_forced_r}) can effectively be approximated with the plane waves
\begin{eqnarray}
\vect A_{\rm forced}(\vect R_0+\vect R,\omega)\approx  
{\rm Re}\left[\vect A_{\rm far}(\vect R_0,\omega)\e^{\i k_0 \vect n\cdot \vect R}\right]
\end{eqnarray} 
of the complex amplitude 
\begin{equation}
	\vect A_{\rm far}(\vect R_0,\omega)=\frac{\mu_0}{2\pi} \frac{\e^{\i k_0\vect n\cdot \vect R_0}}{R_0}
	\int\limits_V\;\vect J(\vect r',\omega)\e^{-\i k_0\vect n\cdot \vect r'}\d^3\vect r',
\end{equation}
where $\vect n$ is the unit vector in the direction of $\vect r$. These plane waves are exactly the free-fields. Hence, for the limited far-field domain we can write
\begin{eqnarray}
\vect A_{\rm forced}(\vect R_0+\vect R,\omega)\approx  
\vect A_{\rm free}(\vect R,\omega).
\end{eqnarray} 
A larger $R_0$ (compared to $r'$, $R$ and $k_0^{-1}$) results in a more accurate fitting provided by the free fields. Then, the relative error of free-field fitting can be estimated with the step function: 0\% for the limited far-field zone and 100\% for the near-field zone. 

\section{Quantum optics limitation} 
\label{sec:Quantum optics limitation}

\begin{figure}[b]
\begin{center}
\begin{tabular}{c}
\includegraphics[trim={0 7cm 0cm 0cm},clip,height=4cm]{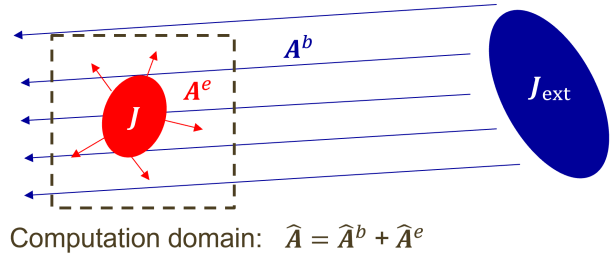}
\end{tabular}
\end{center}
\caption 
{\label{fig:Fig3}
Sketch of bright-field radiation scheme with external current $\vect J_{\rm ext}$ located far away from the computation domain. } 
\end{figure}

Now, we can use the analysis done for the quasi-classical solution, to estimate the error brought by the use of free field to quantum description. For that, we consider a general radiation scheme, the so-called bright-field radiation scheme, with external currents located far away from the computation domain as shown in Fig.~\ref{fig:Fig3}. If we want to include effects of the external currents in our calculations, we can do that through the background fields $\vect A^b(\vect r,t)$ generated by them. Since the computation domain is located far away from the external sources, it can be treated as a limited far-field zone for the background fields. In the quasi-classical limit, these fields are perfectly fitted with the free-field model. We assume that the 0\% fitting error remains valid in the quantum regime of $\hat{\vect A}^b(\vect r,t)$ too. 
As for the fields $\vect A^e(\vect r,t)$ emitted by internal currents within the computation domain, the situation is different. As we already know, the emitted light contains a significant contribution from the near fields ignored in the free-field model. When we apply the free-field model to the emitted light, we bring a significant error. We assume that the 100\% quasi-classical relative error preserves for the operator $\hat{\vect A}^e(\vect r,t)$ in the quantum regime. 

Now, we can estimate the error brought by free-field model to the Hamiltonian operator. As the interaction part of the Hamiltonian operator,
\begin{equation}
\hat H_{\rm int}(t)=\int\hat{\vect J}(\vect r,t)\cdot[\hat{\vect A}^b(\vect r,t)+\hat{\vect A}^e(\vect r,t)]\;\d\vect r,
\end{equation}
is mainly given by the near-field zone of $\hat{\vect J}(\vect r,t)$, then the every emitted photon contributes the near-field-related error to the Hamiltonian operator, while the every background photon is free of it. Based on this, the relative error of quantum calculation for emission of $N^e_{\alpha\vect n}(\omega)$ photons can be estimated with the operator $\hat E(\omega)$ defined as  
\begin{equation}
<N^b_{\alpha\vect n}(\omega)+N^e_{\alpha\vect n}(\omega)|\hat E(\omega)|N^b_{\alpha\vect n}(\omega)>=
\frac{N^e_{\alpha\vect n}(\omega)}{N^b_{\alpha\vect n}(\omega)+N^e_{\alpha\vect n}(\omega)}.\qquad \label{error}
\end{equation}  
To minimize this error, the number of photons $N^b_{\alpha\vect n}(\omega)$ in the background field (or the initial state of light) should be much larger as compared to the number of emitted photons $N^e_{\alpha\vect n}(\omega)$. In the limiting case of dark-field radiation, when we do not have any background field and $N^b_{\alpha\vect n}(\omega)=0$, the interaction part of the Hamiltonian operator appears to contain the 100\% error, making the final results meaningless regardless of the number of emitted photons. 

Highlight that relative error (\ref{error}) is obtained based on the analysis of spin-spin interaction in the quasi-classical limit. So, it can be used as a quasi-classical estimation of the actual error. As for the other types of light-matter interactions, this estimation remains valid, as its mathematical nature is irrespective of the interaction type. This error is brought by the failure of free field to fit forced fields in the near-field zone. Since the free-field model is the only one we have in quantum optics for description of light, error (\ref{error}) naturally appears for every emitted photon regardless of the emission type.  

\section{Conclusions} 
\label{sec:Conclusions}

In conclusion, we highlight the main points of our comparison of classical and quantum optics descriptions. First, quantum optics is unable to provide a closed set of equations for self-consistent description of light-matter interaction. This is why we have to change the light state in quantum calculations independently of the matter state violating the causality principle for light emission. Second, quantum optics does not completely cover the classical optics domain. All near-fields effects appear out of the scope of quantum optics bringing significant errors and restricting quantum calculations to the bright-field radiation problem with sufficiently strong background light.

\begin{backmatter}

\bmsection{Acknowledgments}
The author acknowledges the funding support from Agency for Science, Technology and Research (\#21709).



\end{backmatter}

\bibliography{report}

\begin{thebibliography}{1}
\newcommand{\enquote}[1]{``#1''}

\bibitem{stratton:1941}
J.~A. Stratton, \emph{Electromagnetic Theory} (McGraw-Hill Book Company, 1941).

\bibitem{Landau:1980}
L.~D. Landau and E.~M. Lifshitz, \emph{The classical theory of fields, 4th ed.}
  (Butterworth-Heinemann, 1980).

\bibitem{Landau:1984}
L.~D. Landau and E.~M. Lifshitz, \emph{Electrodynamics of Continuous Media, 2nd
  ed.} (Butterworth-Heinemann, 1984).

\bibitem{Berestetskii:1982}
V.~B. Berestetskii, L.~P. Pitaevskii, and E.~M. Lifshitz, \emph{Quantum
  electrodynamics, 2nd ed.} (Butterworth-Heinemann, 1982).

\bibitem{Feynman:1998}
R.~P. Feynman, \emph{Quantum Electrodynamics} (Westview Press, 1998).

\bibitem{Akimov:2021-1}
Y.~Akimov and P.~Rutkevych, \enquote{Photon model of light: Revision of
  applicability limits,} arXiv:2106.08140 [physics.gen-ph] (2021).

\bibitem{Akimov:2021-2}
Y.~Akimov, \enquote{Light emission problem in classical and quantum optics,} in
  \emph{Education and Training in Optics \& Photonics Conference 2021,}
  (Optica Publishing Group, 2021), p. W2B.7.

\end{thebibliography}

\end{document}